\documentclass[final, 5p, authoryear, times, twocolumn]{elsarticle}

\usepackage{amssymb}
\usepackage{amsmath}
\usepackage{mathrsfs}
\usepackage{booktabs}
\usepackage{color}

\journal{XXXX}

\begin{document}

\begin{frontmatter}



\title{Comparative analysis of layered structures in empirical investor networks and cellphone communication networks}

\author[BS]{Peng Wang}
\author[BS]{Jun-Chao Ma}
\author[BS]{Zhi-Qiang Jiang \corref{cor}} \ead{zqjiang@ecust.edu.cn} %
\author[BS,SS]{Wei-Xing Zhou \corref{cor}} \ead{wxzhou@ecust.edu.cn} %
\author[ETH,SFI]{Didier Sornette} \ead{dsornette@ethz.edu} %
\cortext[cor]{Corresponding authors. Address: 130 Meilong Road, P.O. Box 114, Department of Finance, East China University of Science and Technology, Shanghai 200237, China, Phone: +86 21 64250053, Fax: +86 21 64253152.}

\address[BS]{ School of Business and Research Center for Econophysics, East China University of Science and Technology, Shanghai 200237, China}
\address[SS]{School of Science, East China University of Science and Technology, Shanghai 200237, China}
\address[ETH]{Department of Management, Technology and Economics, ETH Zurich, Scheuchzerstrasse 7, CH-8092 Zurich, Switzerland} %
\address[SFI]{Swiss Finance Institute, c/o University of Geneva, 40 blvd. Du Pont d'Arve, CH-1211 Geneva 4, Switzerland} %

\begin{abstract}
Empirical investor networks (EIN) proposed by \cite{Ozsoylev-Walden-Yavuz-Bildik-2014-RFS} are assumed to capture the information spreading path among investors. Here, we perform a comparative analysis between the EIN and the cellphone communication networks (CN) to test whether EIN is an information exchanging network from the perspective of the layer structures of ego networks. We employ two clustering algorithms ($k$-means algorithm and $H/T$ break algorithm) to detect the layer structures for each node in both networks. We find that the nodes in both networks can be clustered into two groups, one that has a layer structure similar to the theoretical Dunbar Circle corresponding to that the alters in ego networks exhibit a four-layer hierarchical structure with the cumulative number of 5, 15, 50 and 150 from the inner layer to the outer layer, and the other one having an additional inner layer with about 2 alters compared with the Dunbar Circle. We also find that the scale ratios, which are estimated based on the unique parameters in the theoretical model of layer structures \citep{Tamarit-Cuesta-Dunbar-Sanchez-2018-PNAS}, conform to a log-normal distribution for both networks. Our results not only deepen our understanding on the topological structures of EIN, but also provide empirical evidence of the channels of information diffusion among investors.

\end{abstract}


\end{frontmatter}


\section{Introduction}

\cite{Ozsoylev-Walden-Yavuz-Bildik-2014-RFS} proposed the empirical investor network (EIN) as a novel representation of the interactions between investors, based on their order placements: two investors are said to be connected if they placed the same type (ask or bid) of orders within a short time window (usually 30 seconds). The underlying hypothesis behind the EIN is that, when new information comes, it spreads from the source nodes to the peripheral nodes in the investor social networks and the time lags with which the information reaches different investors determine the lags between their order placements. Therefore, EIN can be regarded as a proxy of the investor social network. We propose to check the validity of the EIN construction by studying some of its properties, such as the layer or hierarchical structures in the EIN. As a reference and for comparison, we also test the hierarchical structures present in cellphone communication networks (CN), which are usually considered as information spreading network. Our finding of similar layer structures in EIN and CN gives credence to the hypothesis that EIN uncovers a significant part of the information spreading path between investors.

The present work is related to the research on Dunbar's number and its generalised discrete hierarchical structure in social networks. Recall that Dunbar's number of about 150 represents the average size of the personal ego network, i.e., the group of people one can typically maintain stable social relationships with due to cognitive limits \citep{Dunbar-1992-JHE, Dunbar-1993-BBS}. Furthermore, the social relations in human and animal network have been found to form layer structures, each layer representing different emotional closeness \citep{Dunbar-1998-EAINR, Dunbar-Shultz-2007-Science}. And layer structures have approximately the configuration of 3-5, 10-15, 30-50, and 100-200 alters from the inner layer to outer layer \citep{Zhou-Sornette-Hill-Dunbar-2005-PRSB}. Many empirical ego networks are found to exhibit such layer structures, including the network abstracted from the exchange of Christmas cards \citep{Hill-Dunbar-2003-HN}, the hunter-gatherer social networks \citep{Hamilton-Milne-Walker-Burger-Brown-2007-PRSB, Zhou-Sornette-Hill-Dunbar-2005-PRSB}, and online societies in virtual world \citep{Fuchs-Sornette-Thurner-2014-SR}.

Another strand of literature relevant to our work is the use of 
cellphone and internet communication data that enable one to test the classical social theories empirically in large scale individuals. For example, the weak tie theory \citep{Granovetter-1973-AJS} has been validated for cellphone communication networks \citep{Onnela-Saramaki-Hyvonen-Szabo-Lazer-Kaski-Kertesz-Barabasi-2007-PNAS, Kovanen-Kaski-Kertesz-Saramaki-2013-PNAS}. Such data have also been used to verify the hierarchical layer structures in social networks \citep{Saramaki-Leicht-Lopez-Roberts-ReedTsochas-Dunbar-2014-PNAS}. \cite{Arnaboldi-Dunbar-Passarella-Conti-2016} found that the co-author networks in academic fields also have discrete hierarchical structures. By scanning the online social network from Facebook and Twitter, \cite{Dunbar-Arnaboldi-Conti-Passarella-2015-SN} found that the ego networks exhibit limit size and hierarchical structures. More importantly, such layer structure can be considered as a ``social fingerprint'' for a specific individual, because it is stable and not affected by the change of friends \citep{Tamarit-Cuesta-Dunbar-Sanchez-2018-PNAS}.  

This paper is organized as follows. Data and methods are given in Sec.~\ref{Sec:DataMethods}. Sec.~\ref{Sec:Result} presents the results on the degree distribution, clustering, and theoretical model fits. Sec.~\ref{Sec:Conclusion} concludes. 
	
\section{Data and Methods}
\label{Sec:DataMethods}

\subsection{Empirical investor networks}
	
Our empirical investor networks (EIN) are constructed from the order flows of 100 stocks included in the Shenzhen 100 index (399004). The order flow data span the whole year of 2013. Following \cite{Ozsoylev-Walden-Yavuz-Bildik-2014-RFS}, on each trading day, the EIN is obtained by connecting investors if they submit at least 3 buy (or sell) orders for the same stocks within 30 seconds. By aggregating the EIN on each trading day together, we obtain the annual EIN, which contains 381,345 nodes and 8,143,541 links. \cite{Ozsoylev-Walden-Yavuz-Bildik-2014-RFS} argued that the links in EIN may reflect the potential channels of information diffusion among investors, which could be reveal the
existence of localized structures in social networks formed by investors. Thus, the larger the occurrence of links between two investors, the higher the probability for the existence of social connections between them. We further employ a statistical validated method \citep{Tumminello-Micciche-Lillo-Piilo-Mantegna-2011-PLoS1, Tumminello-Lillo-Piilo-Mantegna-2012-NJP, Li-Jiang-Xie-Micciche-Tumminello-Zhou-Mantegna-2014-SR, Hatzopoulos-Iori-Mantegna-Micciche-Tumminello-2015-QF, Curme-Tumminello-Mantegna-Stanley-Kenett-2015-QF, Gualdi-Cimini-Primicerio-DiClemente-Challet-2016-SR} to check whether two investors are occasionally connected, which provides us with the statistical validated empirical investor networks, abbreviated as SVEIN. 

\subsection{Cellphone communication network}
	
The cellphone call records obtained from one Chinese cellphone operator cover periods from June 28th to July 24th and October 1st to December 31st in 2010. By excluding the days October 12th, November 5th, 6th, 13th, 21st and 27th, and December 6th, 8th, 21st and 22nd on which the data were missing, we have a total of 109 days. In the data, there are 91,911,735 cell phone users and 4,599,472,652 calls. As we cannot access the call records from the other cellphone operators, only the call records in which both mobile phone subscribers belongs to the data provider are included in our analysis, which leads to 1,173,501,607 records. As it is known that the frequency of calls may represent the intimacy between friends, the higher the communication frequency between two cellphone users, the stronger their assumed intimacy.
We exclude the users who are identified as robots, telecom frauds and telephone sales \citep{Jiang-Xie-Li-Podobnik-Zhou-Stanley-2013-PNAS}. Finally, we build cellphone communication networks based on the reciprocal calls between normal users. The statistical validated method mentioned above is also employed to remove the random calls, thus providing us with the statistical validated cellphone communication networks, abbreviated as SVCN. 

\subsection{Statistical validated method}
	
As is well known, EIN and CN contain a great deal of noise: for instance, two investors may submit orders at the same time by 
pure coincidence and callers may make wrong calls to callees. This suggests to remove such irrelevant signals by testing whether 
two nodes are randomly connected. For this, we employ a statistically validated method, proposed by \cite{Tumminello-Micciche-Lillo-Piilo-Mantegna-2011-PLoS1} and used in different systems \citep{Tumminello-Lillo-Piilo-Mantegna-2012-NJP, Li-Jiang-Xie-Micciche-Tumminello-Zhou-Mantegna-2014-SR, Hatzopoulos-Iori-Mantegna-Micciche-Tumminello-2015-QF, Curme-Tumminello-Mantegna-Stanley-Kenett-2015-QF, Gualdi-Cimini-Primicerio-DiClemente-Challet-2016-SR} to extract the links that are not randomly generated. 

For two given nodes $i$ and $j$, the purpose of the statistical validation is to check whether $i$ preferentially connects to $j$. The EIN is taken as an example to illustrate the statistical validation method. Let us denote by $N$ is the total number of transactions between investors in EIN, by $N_{ic}$ the number of transactions initiated by investor $i$, by $N_{jr}$ the number of transactions matched by investor $j$, and by $X = N_{icjr}$ the number of transactions initiated by investor $i$ and matched by investor $j$. We can then calculate the probability of observing $X$ co-occurrences via the following equation 
\citep{Tumminello-Micciche-Lillo-Piilo-Mantegna-2011-PLoS1, Tumminello-Micciche-Lillo-Varho-Piilo-Mantegna-2011-JSM}
\begin{equation}
  \label{EQ:OverExpression}
  H(X|N,N_{ic},N_{jr}) = \frac{C^X_{N_{ic}}C^{N_{jr} - X}_{N - N_{ic}}}{C^{N_{jr}}_N},
\end{equation}
where $C^X_{N_{ic}}$ is a binomial coefficient. We can also estimate the $p$-value associated with the observed $N_{icjr}$ as follows:
\begin{equation}
  \label{EQ:p:Nicjr}
  p(N_{icjr}) = 1 - \sum^{N_{icjr} - 1}_{X = 0} H(X|N,N_{ic},N_{jr}).
\end{equation}
For the EIN, we need to perform $2 \times 8,143,541 = 16,287,082$ tests. The corresponding Bonferroni correction of our multiple testing hypothesis is $p_b = 0.01/N_E$ where $N_E=N(N-1)/2$ is the maximal possible number of edges. If the estimated $p(N_{icjr})$ is less than $p_b$, we can infer that investor $i$ preferentially connects to investor $j$. Otherwise, we conclude that the edge pointed from $i$ to $j$ is randomly generated. 

For a given edge between node $i$ and node $j$ in the CN, we are able to estimate the $p$-value for the number of calls $N_{jcir}$ initiated by $j$ and received by $i$ in a similar way. We need to conduct $2 \times 296,928,030 = 593,856,060$ tests. And the Bonferroni correction is set as $p_b = 0.01/N_E$. When $p(N_{icjr})$ is less than $p_b$, this suggests that individual $i$ preferentially calls individual $j$. Only when the two conditions that (1) $i$ preferentially calls $j$ and (2) $j$ preferentially calls $i$ are simultaneously satisfied, do we conclude that the edge between $i$ and $j$ is significant.

\subsection{Clustering method}

\begin{figure}[htbp]
\centering
\includegraphics[width=0.46\textwidth]{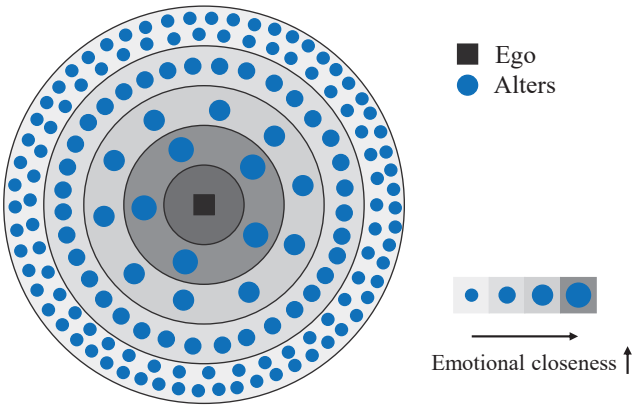}
\caption{Illustration of the theoretical Dunbar Circle in ego networks. The square in the center represent the ego and the circles around are the alters, who have direct connection with the ego. The circle size is proportional to the emotional closeness between the alters and the ego. According to the emotional closeness, the alters form a hierarchical structure with different layers in which their closeness to the ego decrease from inner layer to the outer layer. The theoretical Dunbar Circle corresponds to a four-layer hierarchical structure with the cumulative number of 5, 15, 50, and 150 from inside to outside.} 
\label{Fig:EgoNet:LayerStruct}

\end{figure}

Fig.~\ref{Fig:EgoNet:LayerStruct} illustrates the layer structure of a typical ego network. The ego in the center are surrounded by the alters, who have direct connections with the ego. The alters usually form a layer structure, in which their emotional closeness decrease from the inner layer to the outer layer. The theoretical Dunbar Circle corresponds to a four-layer hierarchical structure with the cumulative number of 5, 15, 50, and 150 from inside to outside. We employ two clustering algorithms, including the $k$-means algorithm and the head-to-tail ($H/T$) break algorithm \citep{Jiang-2013-PG}, to detect the layer structures of the ego network in the SVEIN and SVCN based on the activity frequencies on links. The $k$-means algorithms is implemented with the $R$ package {\bf CKmeans.1d.dp} \citep{Wang-Song-2011-RJ}. The optimized number of clusters are determined according to the BIC. In the $H/T$ break algorithm, the data is split into two parts according to the data mean $m_1$, and the head part in which all values are larger than $m_1$ is further separated into two parts according to the head mean $m_2$. Such process iterates until the head is not heavy-tailed distributed. The $H/T$ break algorithm is proposed to cluster the data with a heavy-tailed distribution, corresponding to the case of link weights in the SVEIN and SVCN.

\section{Result}
\label{Sec:Result}
\subsection{Degree distribution}

We first report the descriptive statistics of both filtered networks. As reported in Panel A of Table~\ref{Tab:StatDesp}, in the SVEIN we find that there are 2.23\%, 6.39\%, and 91.37\% of the total number of users (about 21,806 users) whose degrees are in the range of $k > 100$, $50 < k \le 100$, and $k < 50$, respectively. And their average degree and standard deviation are 142.9 and 38.5, 68.8 and 13.9, and 10.0 and 11.8, leading to a coefficient of variation of 26.95\%, 20.22\%, and  117.95\% (standard deviation/mean). Their average weighted degree and standard deviation are 18487.1 and 10984.6, 5504.3 and 2935.4, and 477.0 and 1134. 

In Panel B of Table~\ref{Tab:StatDesp}, we find that the number of users in the SVCN with degree $k > 100$, $50 < k \le 100$, and $k < 50$ are 60748, 177076, and 3930604, accounting for 1.46\%, 4.25\%, and 94.29\% of the users, respectively. The corresponding average degree and standard deviation are 142.2 and 45.8, 69.4 and 13.7, and 8.1 and 10, resulting in a coefficient of variation of 32.23\%, 19.79\%, 124.08\%. And their average weighted degree and standard deviation are 1544.7 and 775, 780.3 and 410.9, and 92.1 and 161.7. The absolute number of nodes with $k > 100$ in the SVEIN is much smaller than those in the SVCN, but the relative numbers are very close to each other. According to the descriptive statistics, both filtered networks exhibit great similarities in their degree distributions.

\begin{table*}[htbp]
\centering
\caption{Statistical descriptions of SVEIN and SVCN. $k$ denotes the degree of users in the network.}
\begin{tabular}{crrrrrrrr}
\toprule
  &  &  & \multicolumn{3}{c}{Degree} & \multicolumn{3}{c}{Weighted degree}\\
\cmidrule(lr){4-6} \cmidrule(lr){7-9} \  & $N$ & $f$ & mean & std & std/mean & mean & std & std/mean \\
\midrule
Panel A: SVEIN  &       &       &       &       &       &       &       &  \\
$k>100$ & 487 & 2.23\% & 142.9 & 38.5  & 26.95\% &  18487.1  & 10984.6  & 59.42\% \\
$50<k\leq 100$ & 1394 & 6.39\% & 68.8 & 13.9  & 20.22\% &  5504.3  & 2935.4  & 53.33\% \\
$k\leq 50$  & 19925 & 91.37\% & 10.0 & 11.8  & 117.95\% &  477.0  & 1134.0  & 237.73\% \\
\hline
Panel B: SVCN  &       &      &       &       &       &       &       &  \\
$k>100$ & 60748 & 1.46\% & 142.2 & 45.8  & 32.23\% &  1544.7  & 775.0  & 50.17\% \\
$50<k\leq 100$ & 177076 & 4.25\% & 69.4 & 13.7  & 19.79\% &  780.3  & 410.9  & 52.66\% \\
$k\leq 50$  & 3930604 & 94.29\% & 8.1 & 10.0  & 124.08\% &  92.1  & 161.7  & 175.68\% \\
\bottomrule
\end{tabular}
\label{Tab:StatDesp}
\end{table*}

We further fit the empirical degree and weighted degree distributions of the SVEIN and SVCN with the following four distributions, including the power-law, the normal, the exponential, and the log-normal distribution,  
\begin{equation}
f_P(x) = \frac{\alpha-1}{x_{\min}}\left(\frac{x}{x_{\min}}\right)^{-\alpha}, ~~~\alpha >1,
\label{Eq:PDF:PL}
\end{equation}
\begin{equation}
f_N(x) = \frac{1}{\sqrt{2\pi} \sigma_N } \exp \left[ -\frac{(x-\mu_N)^2}{2\sigma_N^2}\right],
\label{Eq:PDF:Norm}
\end{equation}
\begin{equation}
f_E(x) = \lambda e^{-\lambda x}, x>0
\label{Eq:PDF:Exp}
\end{equation}
\begin{equation}
f_L(x) = \frac{1}{\sqrt{2\pi} \sigma_L x} \exp \left[-\frac{(\ln x-\mu_L)^2}{2\sigma_L^2}\right].
\label{Eq:PDF:LogN}
\end{equation}
The parameters of these distributions are obtained by Maximum Likelihood Estimation (MLE). The results are listed in Table~\ref{Tab:PDF:Fits:STests}. Kolmogorov-Smirnov (KS) tests are also conducted to check whether the (weighted) degrees are drawn from the four distributions. The null hypothesis is that the data set follows one of the four distributions. One find that, for both networks, the samples of the degree with $k>0$ and the weighted degree with $k>0$ and $k>100$ conform precisely to none of the four distributions. This is not surprising,
given the large sizes of our data sets, which will thus reject null hypotheses on the basis of even slight deviations.
However, we can still compare the goodness of the fits by the four distributions using the Akaike information criterion (AIC) listed in Table~\ref{Tab:PDF:Fits:STests}. Except for the sample with $k>100$ in the SVEIN, the log-normal distribution has the smallest AIC value. Thus, among the four distributions, the log-normal distribution fits the empirical degree distributions best.

\begin{table*}[htbp]
\small
\centering
\caption{Results of fitting the (weighted) degrees to the power-law, normal, exponential, and log-normal distribution for the SVEIN and SVCN and statistically testing on whether the (weighted) degrees are drawn from the four distributions. The symbols $^{*}$, $^{**}$, and $^{***}$ indicate the significant levels of 5\%, 1\%, and 0.1\%, respectively.}
\begin{tabular}{cr@{.}lr@{.}lr@{.}lr@{.}lr@{.}lr@{.}lr@{.}lr@{.}l}
\toprule
 & \multicolumn{8}{c}{SVEIN} & \multicolumn{8}{c}{SVCN} \\
\cmidrule(lr){2-9} \cmidrule(lr){10-17} \
 & \multicolumn{4}{c}{Degree} & \multicolumn{4}{c}{Weighted degree} & \multicolumn{4}{c}{Degree} & \multicolumn{4}{c}{Weighted degree} \\
\cmidrule(lr){2-5} \cmidrule(lr){6-9} \cmidrule(lr){10-13} \cmidrule(lr){14-17} \
 & \multicolumn{2}{c}{$k>0$} & \multicolumn{2}{c}{$k>100$} & \multicolumn{2}{c}{$k>0$} & \multicolumn{2}{c}{$k>100$} & \multicolumn{2}{c}{$k>0$} & \multicolumn{2}{c}{$k>100$} & \multicolumn{2}{c}{$k>0$} & \multicolumn{2}{c}{$k>100$} \\
\midrule
\multicolumn{8}{l}{Panel A: Fits to the power-law distribution.} \\
$\alpha$ & 1&50  & 3&50  & 1&50  & 1&84  & 1&50  & 3&50  & 1&50  & 1&50  \\
KS    & 0&19  & 0&11  & 0&33  & 0&26  & 0&16  & 0&09  & 0&42  & 0&42  \\
$p$-value   & 0&00$^{***}$  & 0&13  & 0&00$^{***}$  & 0&00$^{***}$  & 0&00$^{***}$  & 0&00$^{***}$  & 0&00$^{***}$  & 0&00$^{***}$  \\
AIC   & 161,511&77  & 4,693&74  & 325,265&95  & 10,568&02  & 28,130,126&25  & 580,173&13  & 49,774,220&93  & 1,084,909&45  \\
\hline
\multicolumn{8}{l}{Panel B: Fits to the normal distribution} \\
$\mu_N$ & -1,031&56  & -93&68  & 1,200&60  & 18,487&09  & 12&61  & -2,292&22  & 142&47  & 1,422&38  \\
$\sigma_N$ & 131&70  & 107&86  & 3,570&82  & 10,973&34  & 23&12  & 327&80  & 298&20  & 881&68  \\
KS    & 0&26  & 0&03  & 0&37  & 0&15  & 0&31  & 0&03  & 0&32  & 0&09  \\
$p$-value   & 0&00$^{***}$  & 0&77  & 0&00$^{***}$  & 0&00$^{***}$  & 0&00$^{***}$  & 0&40  & 0&00$^{***}$  & 0&00$^{***}$  \\
AIC   & 166,010&36  & {\bf 4,632}&{\bf 85}  & 418,656&91  & 10,447&39  & 38,012,462&68  & 576,466&02  & 59,330,827&17  & 975,316&25  \\
\hline
\multicolumn{8}{l}{Panel C: Fits to the exponential distribution} \\
$\alpha$ & 16&71  & 42&91  & 1200&60  & 18487&09  & 12&62  & 42&23  & 142&47  & 1446&69  \\
KS    & 0&24  & 0&05  & 0&41  & 0&28  & 0&24  & 0&02  & 0&30  & 0&28  \\
$p$-value  & 0&00$^{***}$  & 0&46  & 0&00$^{***}$  & 0&00$^{***}$  & 0&00$^{***}$  & 0&63  & 0&00$^{***}$  & 0&00$^{***}$  \\
AIC   & 166,432&62  & 4,637&31  & 352,848&35  & 10,540&11  & 29,472,982&01  & 576,270&11  & 49,680,143&77  & 1,005,464&72  \\
\hline
\multicolumn{8}{l}{Panel D: Fits to the log-normal distribution} \\
$\mu_L$ & 1&83  & 4&65  & 5&08  & 9&67  & 1&61  & 4&20  & 3&34  & 7&24  \\
$\sigma_L$ & 1&43  & 0&39  & 2&04  & 0&54  & 1&30  & 0&54  & 2&04  & 0&45  \\
KS   & 0&13  & 0&04  & 0&06  & 0&08  & 0&11  & 0&02  & 0&07  & 0&01  \\
$p$-value   & 0&00$^{***}$  & 0&71  & 0&00$^{***}$  & 0&00$^{**}$  & 0&00$^{***}$  & 0&61  & 0&00$^{***}$  & 0&17  \\
AIC  & {\bf 157,319}&{\bf 09}  & 4,634&67  & {\bf 314,478}&{\bf 41}  & {\bf 10,208}&{\bf 38}  & {\bf 27,447,309}&{\bf 97}  & {\bf 575,902}&{\bf 17}  & {\bf 45,639,797}&{\bf 91}  & {\bf 955,503}&{\bf 08}  \\
\bottomrule
\end{tabular}
\label{Tab:PDF:Fits:STests}
\end{table*}

\begin{figure}[htbp]
\centering
\includegraphics[width=0.48\textwidth]{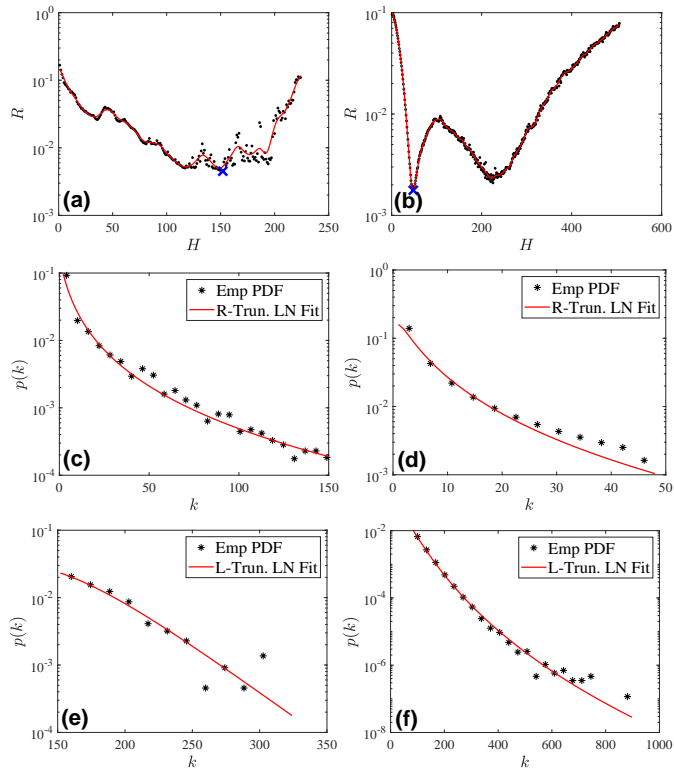}
\caption{Results of the optimal truncated distribution of degrees for SVEIN (a, c, e) and SVCN (b, d, f). (a, b) Plots of the fitting residual (Eq.~(\ref{Eq:PDF:Resi})) as a function of threshold $k_H$. (c, d) Plots of the right-truncated degree distributions.  (e, f) Plots of the left-truncated degree distributions.}
\label{Fig:PDF:Degree}
\end{figure}

\begin{figure}[htbp]
\centering
\includegraphics[width=0.48\textwidth]{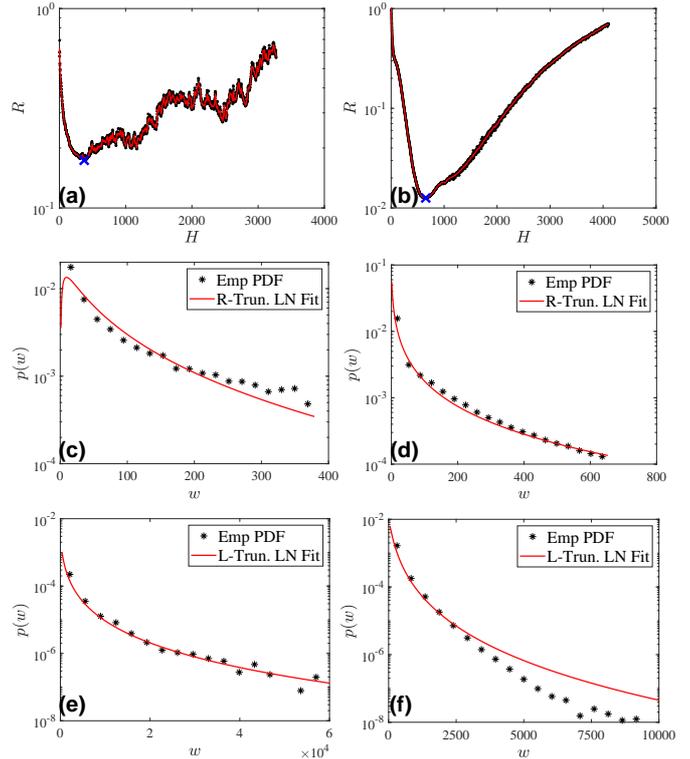}
\caption{Results of the optimal truncated distribution of weighted degrees for SVEIN (a, c, e) and SVCN (b, d, f). (a, b) Plots of the fitting residual (Eq.~(\ref{Eq:PDF:Resi})) as a function of threshold $k_H$. (c, d) Plots of the right-truncated degree distributions.  (e, f) Plots of the left-truncated degree distributions.}
\label{Fig:PDF:WD}
\end{figure}

The results of Table~\ref{Tab:PDF:Fits:STests} strongly suggest that the correct distribution of degrees is a mixture of at least two log-normal distribution, one for small $k$ and one for large $k$. Roughly, we can find a threshold $k_{H}$, the degrees less than $k_{H}$ are fitted by the left truncated log-normal distribution and the degrees greater than $k_{H}$ are fitted by the right truncated log-normal distribution. Following \cite{Wu-Zhou-Xiao-Kurths-Schellnhuber-2010-PNAS,Jiang-Xie-Li-Zhou-Sornette-2016-JSM}, the threshold $k_{H}$ can be estimated by minimizing the following residual, 
\begin{equation}
R = \frac{\left\{\sum_{i}^{n_s}\left[\frac{K_{i,fit}^{s}-K_{i,emp}^{s}}{K_{i,fit}^{s}+K_{i,emp}^{s}}\right] + \sum_{j}^{n_l}\left[\frac{K_{j,fit}^{l}-K_{j,emp}^{l}}{K_{j,fit}^{l}+K_{j,emp}^{l}}\right]\right\}^{\frac{1}{2}}}{\sqrt{n_s+n_l}}
\label{Eq:PDF:Resi}
\end{equation}
where $K_{\rm{fit}}$ and $K_{\rm{emp}}$ represent the fitting distribution and empirical distribution, the superscripts $s$ and $l$ stand for the sample less and greater than the threshold $k_{H}$, and $n$ is the sample size. The parameters of both truncated distributions are determined through the Maximum Likelihood Estimation (MLE). Fig.~\ref{Fig:PDF:Degree} (a) and (b) illustrate the fitting residuals as a function of the possible thresholds for the degrees of SVEIN and SVCN. Thus, we can find that the optimal threshold are 152 and 48 for SVEIN and SVCN, respectively. The corresponding right-truncated and left-truncated degree distributions are plotted in Fig.~\ref{Fig:PDF:Degree} (c -- f) for SVEIN and SVCN. The solid lines in each panel represent the best fits to the truncated log-normal distributions. For the weighted degrees of both networks, we perform the same analysis and illustrate the results in Fig.~\ref{Fig:PDF:WD}. The optimal thresholds are 374 and 653 for the weighted degrees of SVEIN and SVCN, respectively. One can see that the empirical distributions agree well with the fitted distributions in Figs~\ref{Fig:PDF:Degree} and \ref{Fig:PDF:WD}, which support that the (weighted) degrees of both network conform to a mixed log-normal distribution.

As is well known, the log-normal distribution plays an important role in describing natural phenomena in which growth processes are driven by the accumulation of many small percentage changes (growth rates), which is additive on the logarithmic scale. If each percentage change is small enough, the summation on the logarithmic scale tends to be normally distributed according to the central limit theorem, which means that the percentage change follows a log-normal distribution in the linear scale. One intriguing feature of the log-normal distribution is that the growth rate is independent of its size. According to the log-normal degree distributions, one can infer that the growth rate of one's ``friends'' should be independent of one's current number of ``friends'' in the SVEIN and SVCN.

\subsection{Clusters}

\begin{figure}[htbp]
\centering
\includegraphics[width=0.48\textwidth]{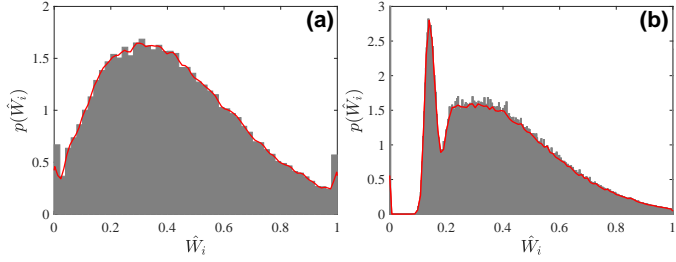}
\caption{Probability distribution of the normalized weights ${\hat{W_i}}$. (a) SVEIN. (b) SVCN.}
\label{Fig:PDF:NormWeights}
\end{figure}

The layer structures in ego networks is usually determined based on the emotional closeness on links. Here, we cannot measure the emotional closeness directly. As an alternative, we employ the number of order placements in the EIN and the number of calls in the CN as a proxy for the emotional closeness on links. For a given node with $n$ links, we first normalize the number of order placements (resp. the number of calls) $W_i$ ($i = 1,~2,~3,~\cdots, n$) on each links via the following equation,
\begin{equation}
\hat{W_i}=\frac{W_{i}-W_{\min}}{W_{\max }-W_{\min}}, 
\label{Eq:Norm:Weights}
\end{equation}
where $W_{\min} = \min(\{W_i\})$ and $W_{\max} = \max(\{W_i\})$. Eq.~(\ref{Eq:Norm:Weights}) insures $0 \le \hat{W_i} \le 1$. 
The presence of natural breaks (associated with network layers) should then be reflected in the existence of sharp peaks in the distributions of $\hat{W_i}$. We thus plot the distribution of the normalized weights $\hat{W_i}$ in Fig.~\ref{Fig:PDF:NormWeights} for both networks. As shown in Fig.~\ref{Fig:PDF:NormWeights} (a), no break can be observed for the SVEIN. A possible explanation is that the data sample of SVEIN is too small. In contrast, there is a significant peak at around $0.1$ for the SVCN, as illustrated in Fig.~\ref{Fig:PDF:NormWeights} (b), which corresponds to the natural break $w_i \approx 0.1 = 15/150$, i.e. the second layer at $15$ of Dunbar's discrete hierarchy.  In the following, we use the clustering algorithm ($k$-means and $H/T$ break) to uncover the discrete hierarchical structure of the node with $k > 100$ based on the normalized weights $\hat{W_i}$.

\begin{figure}[htbp]
\centering
\includegraphics[width=0.48\textwidth]{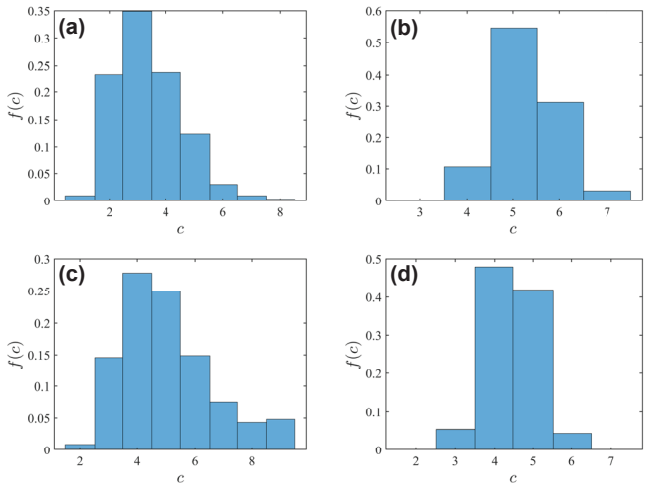}
\caption{Plots of the percentage of the users who have the same number of layers in the SVEIN (a, b) and SVCN (c, d) based on the $k$-means (a, c) and $H/T$ (b, d) break algorithm.}
\label{Fig:Freq:Users:Algo}
\end{figure}

Fig.~\ref{Fig:Freq:Users:Algo} shows the percentage of users who have the same number of layers according to the clustering algorithm of $k$-means and $H/T$ break. As shown in Fig.~\ref{Fig:Freq:Users:Algo} (a) and (b), the alters belonging to investors with degree $k > 100$ in the SVEIN are mainly divided into 2-4 classes and 4-6 classes according to the $k$-means and $H/T$ Break algorithm, respectively. And we also find that 56.9\% of the investors whose alters can be grouped into 5 layers. In order to measure the similarity and robustness of the clustering result, we further estimate the Jaccard coefficient between the clustering results of the two algorithms for the same user. The average Jaccard coefficient of all users is $0.11$. As illustrated in Fig.~\ref{Fig:Freq:Users:Algo} (c) and (d), we find that in the SVCN the alters of the users with degree $k > 100$ are mainly divided into 3-6 classes and 4-5 classes based on the $k$-means algorithm and the $H/T$ Break algorithm. And the average Jaccard coefficient of the clustering results is 0.23. Our results thus indicate that the overlapping of the clusters from both algorithms is low. 

Table~\ref{Tab:Num:Clusters} shows the comparison of the clustering results for the users with degree $k > 100$ in both networks based on the $k$-means and $H/T$ break algorithms. The results of the two clustering algorithms for the SVEIN are reported in panel A of Table~\ref{Tab:Num:Clusters}. We find that 43\% of users with degree $k > 100$ in SVEIN are grouped into 3 layers and the average cumulative number of alters in layers is 10.9, 45.8 and 141.7, in which the last two layers correspond to the middle two layers of the empirical discrete hierarchical structure and the first layer seems to correspond to the coalition of the first two layers of the empirical structure reported previously in \citep{Zhou-Sornette-Hill-Dunbar-2005-PRSB, Hill-Dunbar-2003-HN}. The $H/T$ Break algorithm reveals that about 90\% of the investors whose alters exhibit a configuration with 5 and 6 layers. One can observe that the number of alters in the outer four layers are very close to the theoretical Dunbar Circle 5, 15, 50, and 150. The number of alters in the inner or two layers is only 1-3. 

Panel B of Table~\ref{Tab:Num:Clusters} lists the cumulative number of friends in each layer for the SVCN. For the $k$-means algorithm, we find that 16,918 (a fraction of 41.1\%) users have a four-layer structure. The average cumulative number of alters from inside to outside are 3.0, 12.8, 42.8 and 132.0, which is in agreement with the discrete hierarchical structure 3-5, 10-15, 30-50, and 100-200 previously reported \citep{Zhou-Sornette-Hill-Dunbar-2005-PRSB, Hill-Dunbar-2003-HN}. The corresponding scale ratio is 3.22 which is near to the Dunbar number 3. We also find that there are 15209 users have a five-layer structure with an average accumulative number of 2.1, 7.3, 20.4, 54. 2, and 141.4. Besides the inner layer $n_1 = 2.1$, the number of alters in the outside four layers are very close to the reported hierarchical structure in Ref.~\cite{Zhou-Sornette-Hill-Dunbar-2005-PRSB, Hill-Dunbar-2003-HN}. For the H/T Break algorithm, 29125 users (about 50.2\%) exhibit a four-layer structure and the average cumulative number of alters are 2.1, 8.7, 33.4 and 133.9. There are 25539 (about 44.1\%) users whose alters can be classified into 5 layers and the average accumulative number of alters in successive layers are 1.2, 3.8, 11.7, 39.5 and 147.6. 

Both clustering algorithms reveal a similar discrete hierarchical structure in cellphone networks. We find that there is an extra innermost layer (1.2-2.1), with about 1-2 alters, for the users with four layers in their ego networks. We further fix the number of clusters to 4 for the $k$-means algorithm and estimate the cumulative numbers of in each layer, obtaining 2.5, 10.3, 36.8, and 142.2. 
In addition, we perform the clustering analysis on the link activities for each ego network, in which the ego investor with degrees $50 < k < 100$, by means of the $k$-means algorithm. We find that there are 621 investors (about 44.9\%) having a two-layer structure and the corresponding layer structure is 19.8 and 67.2,  which is close to the middle two layers of the previously reported hierarchical structure \citep{Zhou-Sornette-Hill-Dunbar-2005-PRSB, Hill-Dunbar-2003-HN}. 

The empirical hierarchical structures of the personal ego networks in SVEIN and SVCN are compatible with the structure of 3-5, 10-15, 30-50, 100-200 from the inner to the outer layer, which is close to the theoretical Dunbar Circle. And the average empirical scaling ratio is close to the previously found value 3, which can also be accounted for theoretically \citep{Lera-Sornette-2019-PLoS1}.

\begin{table*}[htbp]
\centering
\caption{Comparison of the clustering results for the users with degree $k >100$ based on the $k$-mean and $H/T$ break algorithm for the SVEIN and SVCN. $N$ and $f$ represents the total number and the percentage of users. $n_k$ stands for the cumulative number of users in the $k$-th layer. $\langle r \rangle$ is the average scale ratio.}
\label{Tab:Num:Clusters}
\begin{tabular}{crrrrrrrrr}
\toprule
 & $N$ & $f$ & $n_1$ & $n_2$ & $n_3$ & $n_4$ & $n_5$ & $n_6$ & $\langle r \rangle$ \\
\midrule
\multicolumn{9}{l}{Panel A: Clustering results of SVEIN} \\
$k$-means &       &       &       &       &       &       &       &  \\
$c$ = 2 & 114 &27.9\% & 27.8 & 121.7 &       &       &       &       & 3.84 \\
$c$ = 3 & 176 &43.0\% & 10.9 & 45.8 & 141.7 &       &       &       & 3.04 \\
$c$ = 4 & 119 &29.1\% & 5.4 & 20.8 & 57.5 & 151.4 &       &       & 2.64 \\
$H/T$ break &       &       &       &       &       &       &       &  \\
$c$ = 4 & 54 &11.3\% & 2.9 & 11.0 & 37.1 & 133.1 &       &       & 3.45 \\
$c$ = 5 & 273 &56.9\% & 1.6 & 5.3 & 15.0 & 42.8 & 133.0 &       & 3.00 \\
$c$ = 6 & 153 &31.9\% & 1.2 & 3.2 & 7.5 & 18.5 & 51.3 & 156.0 & 2.88 \\
\midrule
\multicolumn{9}{l}{Panel B: Clustering results of SVCN} \\
$k$-means &       &       &       &       &       &       &       &  \\
$c$ = 4 & 16918 & 41.1\% & 3.0 & 12.8 & 42.8 & 132.0 &       &       & 3.22 \\
$c$ = 5 & 15209 & 36.9\% & 2.1 & 7.3 & 20.4 & 54.2 & 141.4 &       & 2.66 \\
$c$ = 6 & 9049 & 22.0\% & 1.6 & 5.1 & 12.5 & 28.9 & 66.5 & 154.0 & 2.33 \\
$H/T$ break &       &       &       &       &       &       &       &  \\
$c$ = 3 & 3308 & 5.7\% & 5.0 & 27.1 & 126.7 &       &       &       & 4.71 \\
$c$ = 4 & 29125 &50.2\% & 2.1 & 8.7 & 33.4 & 133.9 &       &       & 3.97 \\
$c$ = 5 & 25539  & 44.1\% & 1.2 & 3.8 & 11.7 & 39.5 & 147.6 &       & 3.61 \\
\midrule
Zhou  & & & 5     & 15    & 50    & 150   &       &       & 3.00 \\
\bottomrule
\end{tabular}
\end{table*}

\begin{figure}[htbp]
\centering
\includegraphics[width=0.46\textwidth]{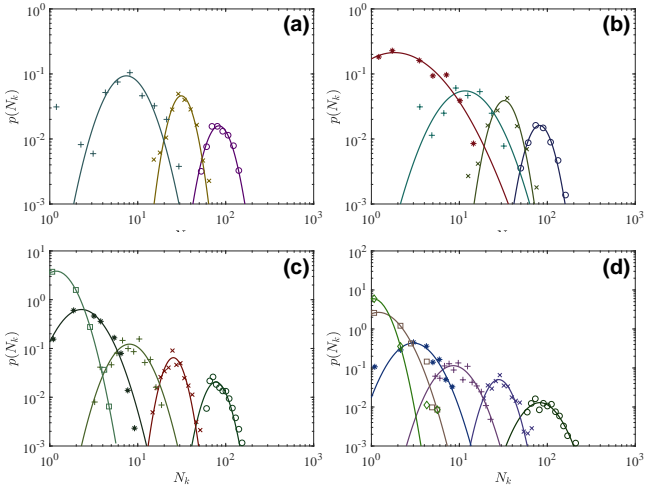}
\caption{Probability distribution of the number of alters in different layers for the SVEIN. The solid curves represent the best log-normal fits to the empirical distribution. (a) Egos with three layers obtained from the $k$-means algorithm. (b) Egos with four layers obtained from the $k$-means algorithm. (c) Egos with five layers obtained from the $H/T$ break algorithm. (d) Egos with six layers obtained from the $H/T$ break algorithm.}
\label{Fig:PDF:LayerUsers:SVEIN}
\end{figure}

\begin{figure}[htbp]
\centering
\includegraphics[width=0.48\textwidth]{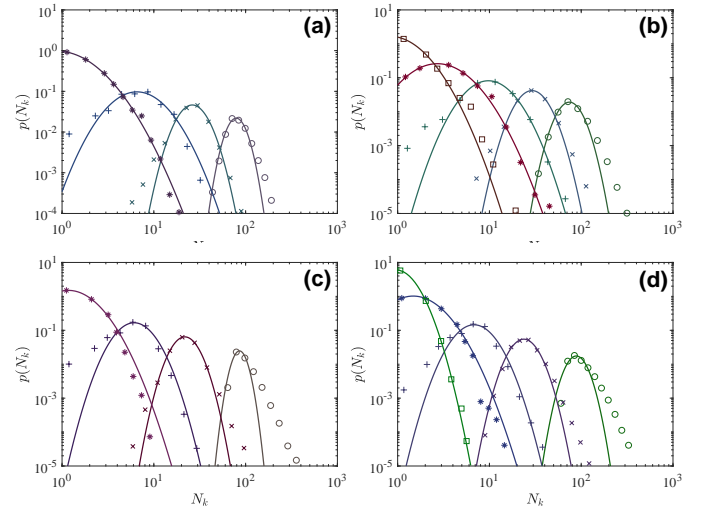}
\caption{Probability distribution of the number of alters in different layers for the SVCN. The solid curves represents the best log-normal fits to the empirical distribution. (a) Egos with four layers obtained from the $k$-means algorithm. (b) Egos with five layers obtained from the $k$-means algorithm. (c) Egos with four layers obtained from the $H/T$ break algorithm. (d) Egos with five layers obtained from the $H/T$ break algorithm. }
\label{Fig:PDF:LayerUsers:SVCN}
\end{figure}

Figs.~\ref{Fig:PDF:LayerUsers:SVEIN} and \ref{Fig:PDF:LayerUsers:SVCN} show the distributions of the numbers of alters in each layer for the egos having degree $k > 100$ in the SVEIN and SVCN. We only show the nodes whose personal ego networks having three-layer and four-layer (respectively, five-layer and six-layer) structures in the SVEIN (SVCN). For both networks, the clustering results of both algorithms are not in agreement with each other, as reflected by the low values of their Jaccard coefficients. An intriguing phenomenon is that the empirical distributions of the number of alters in each layer can be well fitted by the log-normal distributions, evidenced by the solid curves. Such log-normal distribution are robust when using different clustering algorithms, which is in agreement with the results of the online social network from Facebook and Twitter \citep{Dunbar-Arnaboldi-Conti-Passarella-2015-SN}.

\subsection{Fits to the theoretical model}

We further fit the clustering results to the theoretical model of layer structures in personal social network \citep{Tamarit-Cuesta-Dunbar-Sanchez-2018-PNAS}. According to this model, the probability, that the alters of an individual are divided into $\pmb{\ell}=(\ell_1 , \ell_2,...,\ell_r)$, is calculated as follows
\begin{equation}
P(\pmb{\ell} |\mathcal{L},\mu,N) = \mathscr{B}\left(L,\frac{\mathcal{L}}{N-1},N-1\right) \left( \frac{e^{\mu}-1}{e^{\mu r}-1}\right)^L \dbinom{L}{\pmb{\ell}}e^{\mu\sum_ {k=0}^{r-1} k\ell_{k+1} }
\label{Eq:Dunbar:Model}
\end{equation}
where $\pmb{\ell}=(\ell_1,\ell_2,...,\ell_r)$ represents the number of alters in each layer. $\mathcal{L}$ represents the sum of the alters expectation of each layer and is equal to the total number of alters $L$. $N$ is the total number of individuals in the network. $\mathscr{B}(L,p,N) = \dbinom{N}{L}p^L(1-p)^{ NL}$ represents a binomial distribution. There is a unique parameter $\mu$ in the model, which is an indicator of the discrete hierarchy for the ego network. The parameter $\mu$ is approximately equal to the logarithm of the scale ratio $\log(r)$ between the cumulative numbers of individuals in successive layers, if the personal investment (time and energy) decrease linearly with the layers \citep{Tamarit-Cuesta-Dunbar-Sanchez-2018-PNAS}.

\begin{figure}[htbp]
\centering
\includegraphics[width=0.48\textwidth]{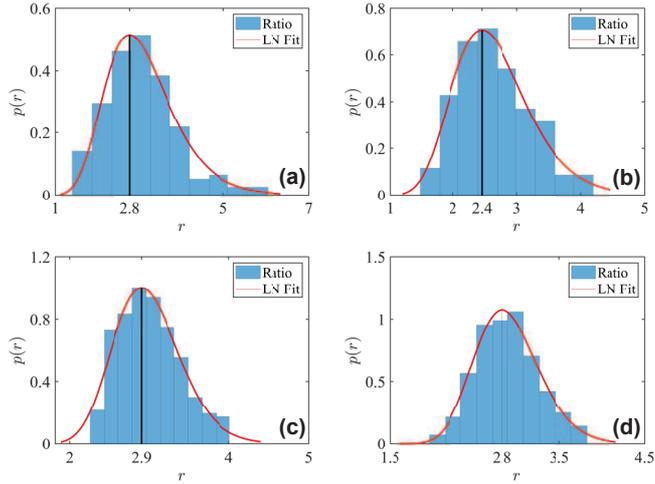}
\caption{Empirical distribution of the scale ratios of the egos with different layers based on different cluster algorithms for the SVEIN. (a) Three layers and $k$-means algorithm. (b) Four layers and $k$-means algorithm. (c) Five layers and $H/T$ break algorithm. (d) Six layers and $H/T$ break algorithm.}
\label{Fig:PDF:SR:SVEIN}
\end{figure}	

Once the empirical hierarchical structure of egos is obtained, we calculate the average scale ratio $\langle r \rangle$ between adjacent cumulative layers based on the model proposed by Tamarit \citep{Tamarit-Cuesta-Dunbar-Sanchez-2018-PNAS}. The estimated theoretical scale ratios of both algorithms are listed in the last column of Table~\ref{Tab:Num:Clusters}. For the SVEIN, the $k$-means algorithm indicates that the users are preferentially divided into the group having a three-layer structure while the $H/T$ break algorithm uncovers that the ego networks exhibit a configuration of five layers. And their scale ratio are very close to the scaling ratio $3$ discovered by \cite{Zhou-Sornette-Hill-Dunbar-2005-PRSB}. However, we find the existence of significant differences in the average scale ratio between the two clustering algorithms for the SVCN. On average, the average scale ratio of the $H/T$ break algorithm is larger than 3.5 and the scale ratio obtained with the $k$-means algorithm is smaller than 3.5. Both clustering algorithms reveal that most of the users exhibit a four-layer structure in their ego networks, for which the scale ratio are respectively 3.2 and 4.0, which are roughly compatible with the scale ratio reported previously \citep{Zhou-Sornette-Hill-Dunbar-2005-PRSB}. 

\begin{figure}[htbp]
\centering
\includegraphics[width=0.48\textwidth]{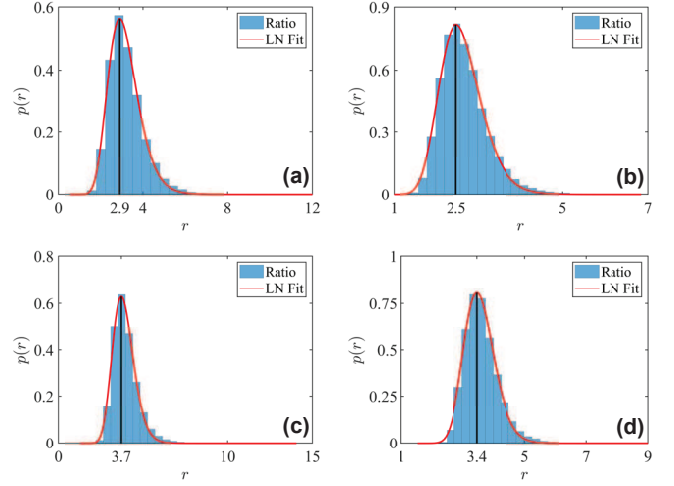}
\caption{Empirical distribution of the scale ratios of the egos with different layers based on different cluster algorithms for the SVCN. (a) Four layers and $k$-means algorithm. (b) Five layers and $k$-means algorithm. (c) Four layers and $H/T$ break algorithm. (d) Five layers and $H/T$ break algorithm.}
\label{Fig:PDF:SR:SVCN}
\end{figure}

Figs.~\ref{Fig:PDF:SR:SVEIN} and \ref{Fig:PDF:SR:SVCN} show the distribution of the estimated average scale ratios for the egos having the same layer structure for both networks. We find that the scale ratio distributions given by the Tamarit's model conform to the log-normal distributions for both clustering algorithms. The $\chi^2$ test, KS test and AD test can not reject the null hypothesis, that the scale ratio are log-normal distributed, at the significant level of 5\%. The solid curves in Figs.~\ref{Fig:PDF:SR:SVEIN} and \ref{Fig:PDF:SR:SVCN} are the best fits to the log-normal distributions. The estimated $\hat{\mu}$ of the scaling ratios are located in the range of 2.5-3.3, which is compatible with the previous scaling ratio $3$ discovered by  \cite{Zhou-Sornette-Hill-Dunbar-2005-PRSB}. Our results reveal that the ego networks in SVEIN exhibit very similar layer structures to those in SVCN, confirming that the SVEIN captures the information spreading channels between investors.

\section{Conclusion}
\label{Sec:Conclusion}
We have performed a comparative analysis to detect the layer structures in Empirical Investor Networks and Cellphone Communication Networks. The layer structures have been quantified by two clustering algorithms, namely the $k$-means and $H/T$ break algorithms. And both clustering algorithms reveal that there are two types of inner structure for both networks: one exhibits a layer structure similar to that of the theoretical Dunbar Circle, while the other has an additional inner layer, which is also found in Facebook and Twitter datasets \cite{Dunbar-Arnaboldi-Conti-Passarella-2015-SN}. Furthermore, we also find that both networks have a similar scale ratio (close to 3). And more interesting, these scale ratios remain stable even when old alters are replaced by new alters. By fitting our empirical clustering results to the theoretical model of layer structures \citep{Tamarit-Cuesta-Dunbar-Sanchez-2018-PNAS}, we confirm that the scale ratios of different egos follow a log-normal distribution for both networks. Our results suggest strong evidence that the structures of ego networks in EIN and CN exhibit great similarities, which captures the information spreading routes between investors and validates the underlying assumption of EIN. 

\bigskip
{\textbf{Acknowledgments:}}

This work was partially supported by the National Natural Science Foundation of China (91746108),  the Shanghai Philosophy and Social Science Fund Project (2017BJB006), the Program of Shanghai Young Top-notch Talent (2018), and the Fundamental Research Funds for the Central Universities. 

\bibliography{E:/Papers/Auxiliary/Bibliography}

\end{document}